\documentstyle[11pt]{l-aa}

\begin{document}
\voffset=-0.6 in

\title{Gamma-ray burst afterglows from jetted shocks in wind environments}

\author{ L.J.~Gou\inst{1}, Z.G.~Dai\inst{1,2}, Y.F.~Huang\inst{1,3}, and T.~Lu\inst{1,2} }

\offprints{ T.~Lu (E-mail: tlu@nju.edu.cn)  Z.G.~Dai (daizigao@public1.ptt.js.cn)  }

\institute{ Department of Astronomy, Nanjing University, Nanjing 210093,
            P. R. China
\and         
            IHEP, Chinese Academy of Sciences, Beijing 100039, P.R. China
\and
            Astronomical and Astrophysical Center of East China,
            Nanjing University, Nanjing 210093, P. R. China
            }
            
%\institute{ Department of Astronomy, Nanjing University, Nanjing 210093, 
%	    P.R. China
%\and      
%            IHEP, Chinese Academy of Sciences, Beijing 100039, P. R. China
%            
%\and       
%            Astronomical and Astrophysical Center of East China, 
%            Nanjing University, Nanjing 210093, P. R. China
%          }

\thesaurus{13.07.1, 09.10.1, 08.13.2, 02.19.1}

\date{Received  6 June 2000; accepted 22 September 2000}

\maketitle
\markboth{L.J. Gou et al.: Jets in GRBs}{}

\begin{abstract}
Gamma-ray bursts  with long durations are widely thought to arise from the  collapse of massive stars, 
where the wind environment is unavoidable. 
It is also believed that $\gamma$-ray bursts come from jets. Considering these two points in this paper, we calculate the evolution 
of a highly collimated jet that expands in a stellar wind environment and the expected afterglow from such a jet. 
 We use a set of refined dynamical equations and a realistic lateral speed of the jet, and find: 
 (1) There is no observable break at the
time when the  Lorentz factor of the jet is equal to the inverse of its initial half-opening angle. 
(2) No obvious break appears at the time when the blast wave transits from the relativistic to the non-relativistic phase. 
 (3) For the wind case, there is no flattening tendency even up to $10^9$ s.
  (4) Compared with the homogeneous medium case, our calculated  flux is weaker in the stellar wind case.
Finally, we find that two kinds of  GRB models (neutron star mergers and massive star collapses) may be discriminated  in our numerical results.


\end{abstract}

\section{Introduction}

Since the discovery of the GRB 970228 afterglow by BeppoSAX, research on gamma-ray bursts (GRBs) has evolved. 
Now we know that GRBs are one of the most energetic phenomena at the cosmological distance. 
Optical afterglows have been observed from about a dozen GRBs (Klose 2000), most of which concentrate at the 
distance scale of $z\sim 1$, corresponding to the luminosity distance of about 3.0 Gpc. In addition, some GRBs' host galaxies 
have been discovered. All these discoveries  leave no doubt that GRBs, the origins of which had puzzled people since their
discovery more than 30 years ago, are of cosmological origin.

Whether a jet exists or not in GRBs is a fundamental problem. We believe that 
jets should exist based on the following facts: (1) The isotropic energy release per GRB is generally in the range of $10^{51} - 10^{52}$ ergs. 
It can be explained well by a stellar-mass progenitor. However for two GRBs, GRB 990123 (Kulkarni et al. 1999a) and  GRB 990510 (Harrison et al. 1999), 
the isotropic energy is so enormous that it is difficult  to explain it by any stellar progenitor model, which forces some theorists to deduce that 
the radiation must be highly collimated in these cases. (2) The steepening of some afterglow light curves observed at the optical band  is argued as 
evidence that jets exist in GRB radiation (Kulkarni et al. 1999a; Harrison et al. 1999; Stanek et al. 1999; Huang et al. 2000a).
Rhoads (1997a, 1999), Sari, Piran  \& Halpern (1999), M\'{e}sz\'{a}ros \& Rees (1999b) have shown that the lateral 
expansion of a relativistic jet will lead to a more rapid deceleration, causing a sharp  
break in the afterglow light curve. For GRB 990123, the power law index of the afterglow light curve is $\alpha = 1.0 \pm$
0.03 in the 2 days after the burst. After 2 days, there is a sudden steepening in the light curve (Kulkarni et al. 1999b). Similarly, for GRB 990510 
the power law index changes from $\alpha$ = 0.76 to $\alpha = 2.4 \pm$ 0.02 after t = 1.0 day (Stanek et al. 1999). Recently, a rapid decay 
with $\alpha =$ 1.73 was found in GRB 970228 (Galama et al. 2000). All these breaks may be due to the lateral expansion of jets (Huang et al. 2000a). 
(3) The observed radio flare may provide an independent and excellent indication of a jet-like geometry in 
GRBs (Harrison et al. 1999;  Kulkarni et al. 1999).  As argued by Waxman, Kulkarni and Frail (1998), the radio afterglow from a spherical fireball 
must rise to a peak flux on a timescale of a few weeks, but  because of lateral expansion, the radio afterglow from the forward shock of a 
jet must fade a few days after the burst. Therefore, the relative faintness of the observed late-time radio 
emission implies the existence of a jet (Sari \& Piran 1998a). GRB 990510 may be a good example: its radio radiation began to decline one day after
the burst (Harrison et al. 1999). (4) The polarization observed in the afterglows may also be evidence for jets. Gruzinov (1999) has
argued that optical afterglows from jets can be strongly polarized, in principle up to tens of percents, if co-moving 
magnetic fields parallel and perpendicular to the jet have different strengths and if we observe the afterglows at a right viewing 
angle.  However, Sari (1999) argued that even if the magnetic fields had a well-defined orientation relative to the direction of 
the shock, the polarization is unlikely to exceed 20\%. Furthermore, taking into account the dynamics of jets,
the polarization first rises to the peak around the jet break time and then decays. (5) Observed light curves of some GRBs steepen 
simultaneously at  different bands. This may be further evidence that there is
collimated ejecta in GRB radiation (Harrison et al. 1999). (6) Observational characteristics of  some GRBs are similar to those of 
BL Lacs, in which jets are unavoidably involved (Paczy\'{n}ski 1993; Dermer \& Chiang 1999; Cheng, Fan \& Dai 1999).
This implies that GRBs may arise from jets.
      
Generally, the broken light curves of some afterglows are explained to be due to the expansion of jets  into a homogeneous interstellar 
medium (Harrison et al. 1999; Stanek et al. 1999; Huang et al. 2000a). However,  other explanations have also been proposed. For example, when 
a spherical fireball evolves in the Wolf-Rayet star wind, the light curve can also  steepen  (Chevalier \& Li 1999, 2000; M\'{e}sz\'{a}ros, Rees 
\&  Wijers 1998; Frail et al. 1999a, b; Dai \& Lu 1998a). Dai \& Lu (1999, 2000a, b) suggested that when a shock in a dense medium transits
 from the relativistic phase to the non-relativistic phase, a break  would occur in the light curve.

The most important problem in GRB research is the energy mechanism. It is widely believed that 
GRBs with long durations come from the  collapse of massive stars (Fryer et al. 1998, 1999a, b; In't Zand 1998;
Ruffert \& Janka 1998, 1999; Woosley, MacFadyen  \& Heger 1999; Rees  1999; M\'{e}sz\'{a}ros \& Rees 1999a). For a massive star at the end of 
its evolution, it throws away its envelope and the core collapses into a compact object, producing a jet. Then due to 
interactions between different shells inside the jet and between the jet and its surrounding medium, a GRB and its 
afterglow are produced, respectively. The recently discovered connection between supernovae and GRBs provides a 
strong support to such a collapsar model. At present, there are three 
GRBs which are most likely connected with supernovae: GRB 980425 (SN 1998bw) (Galama et al. 1998; Iwamoto et al. 1998), 
GRB 970228 (Galama et al. 2000; Reichart  1999), GRB 980326 (Bloom et al. 1999). Recently, 
two other GRBs were added, i.e. GRB 970514 (SN 1997cy) (Germany et al. 2000; Turatto, Suzuki \& Mazzali 2000) and GRB 980910a (SN 1999e) 
(Kulkarni \& Frail 1999c; Thorsett \& Hogg 1999). 

Livio \& Waxman (1999) recently discussed the evolution of a jet in the wind environment  and gave 
an analytical result. They argued that at late stages (particularly after the break corresponding to $\gamma=1/\theta_{\rm o}$), 
the light curve has a  flattening tendency. 
In this paper, we use some refined equations to describe the evolution of jets. First, in the adiabatic case, when blast 
wave is  extremely  relativistic, its dynamical evolution satisfies the Blandford-Mckee (1976) solution. But when it 
reaches the non-relativistic phase, it satisfies the Sedov-Taylor self-similar solution. However, the conventional dynamical model can not 
transit correctly from the ultra-relativistic phase to the non-relativistic phase. This has been stressed by Huang et al. (1998a, b, 1999a, b). 
Here we use the refined dynamical equations proposed by Huang et al. (1999a, b, 2000b, c, d), which can describe the overall evolution of 
jets from the ultra-relativistic phase to the non-relativistic phase. Second, for the lateral expansion speed of jets, it is reasonable to assume 
that it is just the co-moving local sound speed $c_{\rm s}$. Usually, one has taken $c_{\rm s}=c$ or $c/\sqrt{3}$ (Rhoads 1997a, b, 1999; 
Sari, Piran \& Halpern 1999),  where $c$ is the speed of light. In fact we expect $c_{\rm s}$ to 
vary with time, and especially it will by no means be $c$ or $c/\sqrt{3}$  when the blast wave decelerates into the 
non-relativistic 
stage. Huang et al. (2000b, c, d) have given the proper lateral expansion speed which depends on the bulk speed of the blast wave.

Based on these considerations, we calculate the evolution of jets in the wind environment from the relativistic stage to 
the non-relativistic stage and compare numerical results with those of jet evolution in the homogeneous medium. 
We describe our model in section 2. Our detailed numerical results are 
presented in section 3. Section 4 is a brief discussion of our final results.

\section{Model}

\subsection{Basic Equations}

The overall evolution of a jet can be described by the refined equations (Huang et al. 2000b, c, d)

\begin{eqnarray}
\label{drdtb1}
\frac{d \gamma}{d t_{\oplus}}&=&-\frac{(\gamma^2-1)}{M_{\rm ej}+\epsilon m_{\rm sw}+2(1-\epsilon)\gamma m_{\rm sw}} \nonumber \\
      &&  \times  2\pi R^2(1-\cos \theta_{\rm j})\rho \beta  c\gamma(\gamma+\sqrt{\gamma^2-1}),
\end{eqnarray}

\begin{equation}
\label{dtbdt2}
\frac{dm_{\rm sw}}{dt_{\oplus}} = 2\pi R^2 \left(1-\cos \theta_{\rm j}\right)\rho \beta c\gamma \left(\gamma+\sqrt{\gamma^2-1}\right),  
\end{equation}

\begin{equation}
\label{drdt3}
\frac{d R}{d t_{\oplus}} = \beta c \gamma \left(\gamma + \sqrt{\gamma^2 - 1}\right),
\end{equation}

\begin{equation}
\label{dmdr4}
\frac{d \theta_{\rm j}}{d t_{\oplus}} = \frac{1}{R}c_{\rm s} \left(\gamma+\sqrt{\gamma^2-1} \right),
\end{equation}
where $\gamma$ is the Lorentz factor of the jet, $m_{\rm sw}$ is the total mass of the swept-up medium, $R$ is the radius, $\theta_{\rm j}$ is
 the half opening angle of the jet, $t_{\oplus}$ 
s the observed time,  $\rho$ is the mass density of the interstellar 
medium, $M_{\rm ej}$ is the initial mass of the jet, and $\epsilon$ is the radiation efficiency of the jet.
Equations (1) $-$ (4) describe the overall dynamical 
evolution. However before evaluating them numerically, we should give the expressions for $c_{\rm s}, \epsilon$, and  $\rho$.

\subsection{Sound Speed}

The lateral expansion is determined by the co-moving sound speed. 
The simple assumption of $c_{\rm s} = c/\sqrt{3}$ is unreasonable in the present paper. 
Huang et al. (2000b, c, d) give  the proper sound speed which depends on the bulk Lorentz factor. Here we give a brief derivation.
Kirk \& Duffy (1999) have derived 
\begin{equation}
\label{kirkcs5}
c_{\rm s}^2 = \frac{\hat{\gamma} p'}{\rho '} 
	      \left[ \frac{(\hat{\gamma}-1) \rho '}
			   {(\hat{\gamma}-1) \rho ' + \hat{\gamma} p'} 
  	      \right] c^2 ,
\end{equation}
where $\rho '$ and $p'$ are the co-moving mass density and pressure, respectively, 
and $\hat{\gamma}$ is the adiabatic index. Dai, Huang \& Lu (1998, 1999)  
obtained a simple and useful expression for $\hat{\gamma}$, 
$\hat{\gamma} = \left(4 \gamma + 1 \right) / \left(3 \gamma \right)$.  
Since $e' = \gamma \rho ' c^2$ 
and $p' = (\hat{\gamma}-1)(e' - \rho ' c^2)$, 
it is easy to get (Huang et al. 2000b, c, d)
\begin{equation}
\label{cssquare6}
c_{\rm s}^2 = \hat{\gamma} (\hat{\gamma} - 1) (\gamma - 1) 
	      \frac{1}{1 + \hat{\gamma}(\gamma - 1)} c^2 . 
\end{equation}

In the ultra-relativistic limit ($\gamma \gg 1, \hat{\gamma} \approx 4/3$), 
equation~(\ref{cssquare6}) becomes $c_{\rm s}^2 = c^2/3$; and in the 
non-relativistic limit ($\gamma \sim 1, \hat{\gamma} \approx 5/3$), we 
simply get $c_{\rm s}^2 = 5 \beta^2 c^2/9$. So, equation~(\ref{cssquare6}) 
is a reasonable expression and will be used in our model.

\subsection{Radiative Efficiency}

As usual we assume that the magnetic energy density in the co-moving 
frame is a fraction $\xi_{\rm B}^2$ of the total thermal energy density 
(Dai et al. 1998, 1999)
\begin{equation}
\label{bsquare7}
\frac{B'^2}{8 \pi} = \xi_{\rm B}^2  \frac{\hat{\gamma}\gamma + 1}
		    {\hat{\gamma} - 1} (\gamma - 1) n m_{\rm p} c^2. 
\end{equation}
 Generally, the afterglow comes from synchrotron radiation of the electrons 
accelerated behind the shock (Sari et al. 1998; Wijers, Rees, M\'{e}sz\'{a}ros 1997). 
 The contribution of the inverse 
 Compton-scattering emission is always neglected, because it is unimportant particularly at late 
 times (Waxman 1997). So  we only consider synchrotron emission from the electrons.
   Assume that the accelerated electrons carry a fraction $\xi_{\rm e}$ of the proton energy. This implies that 
   the minimum Lorentz factor of the random motion of electrons in the co-moving frame is

\begin{equation}
\label{gemin8}
\gamma_{\rm e,min} = \xi_{\rm e} (\gamma - 1) \left(\frac{m_{\rm p}}{m_{\rm e}}\right)
		     \left(\frac{{p} - 2}{{ p} - 1}\right) + 1,
\end{equation}
where $m_{\rm e}$ is the electron mass, ${p}$ is the index characterizing 
the power law energy distribution of electrons, and $m_{\rm p}$  is the proton mass.
Considering only the synchrotron radiation, Dai et al.(1998,1999) derived the radiative efficiency of the jet

\begin{equation}
\label{radeps9}
\epsilon=\xi_{\rm e} \frac{t^{\prime -1}_{\rm syn}}{t^{\prime -1}_{\rm syn}
	 +t^{\prime -1}_{\rm ex}} ,
\end{equation}
where  $t^{\prime}_{\rm syn}=6\pi m_{\rm e}c/(\sigma_{\rm T}B^{\prime 2}\gamma_{\rm e,min})$ is the synchrotron cooling 
time with   $\sigma_{\rm T}$    being the Thompson scattering cross section, and  $t^{\prime}_{\rm ex}=R/ \gamma c$ is 
the co-moving frame expansion time. For the highly radiative expansion, $\xi_{\rm e} \approx 1$ 
and $t^{\prime}_{\rm syn} \ll t^{\prime}_{\rm ex}$  we have $\epsilon \approx 1$. The early evolution of the jet 
might be in this regime. For the adiabatic expansion, $\xi_{\rm e} \ll 1$ 
or $t^{\prime}_{\rm syn} \gg t^{\prime}_{\rm ex}$, we get $\epsilon \approx 0$. The late evolution  of the jet 
 is believed to be in this regime. In this paper, we consider the latter case, i.e. adiabatic expansion. 

\subsection{Mass Density}

Huang et al. (2000b, c, d) have considered the case that a jet evolves in the homogeneous interstellar medium (ISM). In this paper, we consider 
the case that a jet expands in the preburst stellar wind. For massive stars, particularly Wolf-Rayet stars, the 
typical wind-loss rate is  \( \dot{M} \approx 10^{-5} - 10^{-4}M_{\odot} {\rm yr}^{-1} \), and their typical speed  is  
$V_{\rm w} \approx 1000 $-$ 25000 {\rm km\cdot s}^{-1}$ (Willis 1991). According to $4\pi R^2\rho V_{\rm w}=\dot{M} $, 
we can easily get 

\begin{eqnarray}
\label{radeps10}
\rho&=&5.02\times10^{-18}{\rm g} \cdot {\rm cm}^{-3} \left(\frac{R}{10^{15}\,{ \rm cm}} \right)^{-2}\nonumber \\
&& \times \left( \frac{\dot{M}}{10^{-4}M_{\odot} { \rm yr}^{-1}}\right)\left(\frac{V_{\rm w}}{1000\,{\rm km}\cdot {\rm s}^{-1}}\right).
\end{eqnarray}
 We will use this kind of mass density structure in the paper.

\begin{figure}
\begin{picture}(100,170)
\put(0,0){\includegraphics{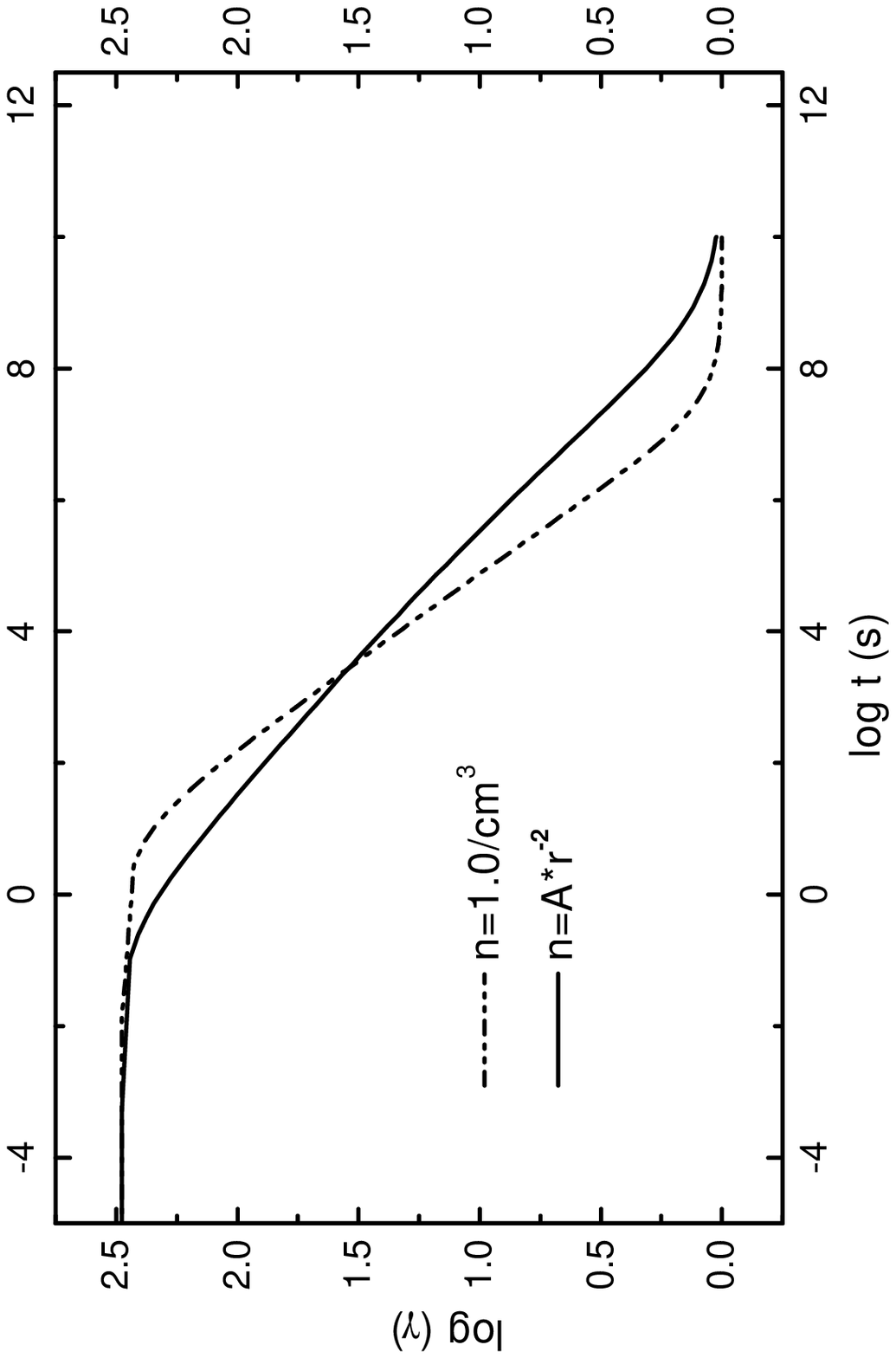}}
\end{picture}
\caption
{ Evolution of the Lorentz factor. We have taken the ``standard'' parameters: $E_{\rm 0}/\Omega_{\rm 0}
=1 \times 10^{54} {\rm ergs}/4\pi, {\rm \gamma}_{\rm 0}=300$ (i.e. $M_{\rm ej}/\Omega_{\rm 0}=0.002M_\odot /4\pi$), $\xi^{2}_{\rm B}=0.02$, $p=2.5$,
${\rm \xi}_{\rm e}=0.1,  {\rm \theta}_{\rm 0}=0.2$. The solid line corresponds to the wind environment ($\dot{M}=
10^{-5}M_\odot\,{\rm yr}^{-1}$ and $V_{\rm w}=10^3\,{\rm km}\,{\rm s}^{-1}$), and the dashed line  to the homogeneous ISM
($n=1\,{\rm cm}^{-3}$). The time that $ \gamma $ = 2 is  about 6 $\times 10^{6}$ s in the ISM case, and  $10^{8}$ s in the wind case. }
\end{figure}

\subsection{Electron Energy Distribution}

In the absence of radiation loss, the distribution of the shock accelerated electrons behind the blast wave is usually assumed to be a power law function 
of electron energy, 
\begin{equation}
\label{dndr11}
\frac{dN_{\rm e}'}{d\gamma_{\rm e}} \propto \gamma_{\rm e}^{-{p}},
\,\,\,\,\,\,(\gamma_{\rm e,min}\leq \gamma_{\rm e} \leq\gamma_{\rm e,max}),
\end{equation}
where $\gamma_{\rm e,max}$ is the maximum Lorentz factor, 
$\gamma_{\rm e,max}=10^8(B^{\prime}/1{\rm G})^{-1/2}$ (Dai et al. 1998, 1999), 
and ${ p}$ usually varies between 2 and 3. However, radiation 
loss may play an important role in the process. 
Electrons with different Lorentz
factors have different radiation efficiencies.
Sari, Piran \& Narayan (1998) have derived an equation for 
the critical electron Lorentz factor, $\gamma_{\rm c}$,
above which synchrotron radiation is significant,
\begin{equation}
\label{gammac12}
\gamma_{\rm c}=\frac{6 \pi m_{\rm e} c}{\sigma_{\rm T} \gamma B'^2 t}.
\end{equation}
Electrons with Lorentz factors below $\gamma_{\rm c}$ are 
adiabatic, and electrons above $\gamma_{\rm c}$ are highly radiative.

In the presence of a steady injection of electrons accelerated by the shock,
the distribution of radiative electrons becomes another power law
function with an index of ${p}+1$ (Rybicki \& Lightman 1979), but the
distribution of adiabatic electrons is unchanged. Then the actual
distribution should be given according to the following cases 
(Dai et al. 1998d, 1999):

\begin{description}
\item (1) For $\gamma_{\rm c}\leq \gamma_{\rm e,min}$,
\begin{equation}
\label{dnei13}
\frac{dN_{\rm e}'}{d\gamma_{\rm e}}=C_1\gamma_{\rm e}^{-({p}+1)}\,, \,\,\,\,\,\,
(\gamma_{\rm e,min}\leq\gamma_{\rm e}\leq \gamma_{\rm e,max})\,,
\end{equation}
\begin{equation}
\label{dneic14}
C_1=\frac{p}{\gamma_{\rm e,min}^{-{p}}- \gamma_{\rm e,max}^{-{p}}}N_{\rm ele}\,, 
\end{equation}
where $N_{\rm ele}$ is the number of radiating electrons in a ring between
$\theta$ and $\theta+d\theta$ with $\theta$ being the angle between the velocity 
of emitting material and the line of sight.

\item (2) For $\gamma_{\rm e,min} < \gamma_{\rm c} \leq \gamma_{\rm e,max}$,
\begin{equation}
\label{dneii15}
  \frac{dN_{\rm e}'}{d\gamma_{\rm e}} = \left \{
   \begin{array}{ll}
 C_2\gamma_{\rm e}^{-{p}}\,, \,\,\,\, & (\gamma_{\rm e,min} \leq \gamma_{\rm e}
					      \leq \gamma_{\rm c}), \\
 C_3\gamma_{\rm e}^{-({p}+1)}\,, \,\,\,\, & (\gamma_{\rm c}<\gamma_{\rm e}
					      \leq \gamma_{\rm e,max}),
   \end{array}
   \right. 
\end{equation}
where 
\begin{equation}
\label{dneiic16}
 C_2=C_3/\gamma_{\rm c}\,,
\end{equation}
\begin{equation}
\label{dneiicc17}
 C_3=\left[\frac{\gamma_{\rm e,min}^{1-{p}}-\gamma_{\rm c}^{1-{p}}}
      {\gamma_{\rm c}({p}-1)}+ \frac{\gamma_{\rm c}^{-{p}}-\gamma_{\rm e,max}^{-{p}}}
      {{p}} \right]^{-1}N_{\rm ele} .
\end{equation}

\item (3) For $\gamma_{\rm c} > \gamma_{\rm e,max}$, 
\begin{equation}
\label{dneiii18}
\frac{dN_{\rm e}'}{d\gamma_{\rm e}}=C_4\gamma_{\rm e}^{-{p}}, \,\,\,\,\,\,
(\gamma_{\rm e,min}\leq\gamma_{\rm e}\leq\gamma_{\rm e,max}),
\end{equation}

\begin{figure}
\begin{picture}(100,170)
\put(0,0){\includegraphics{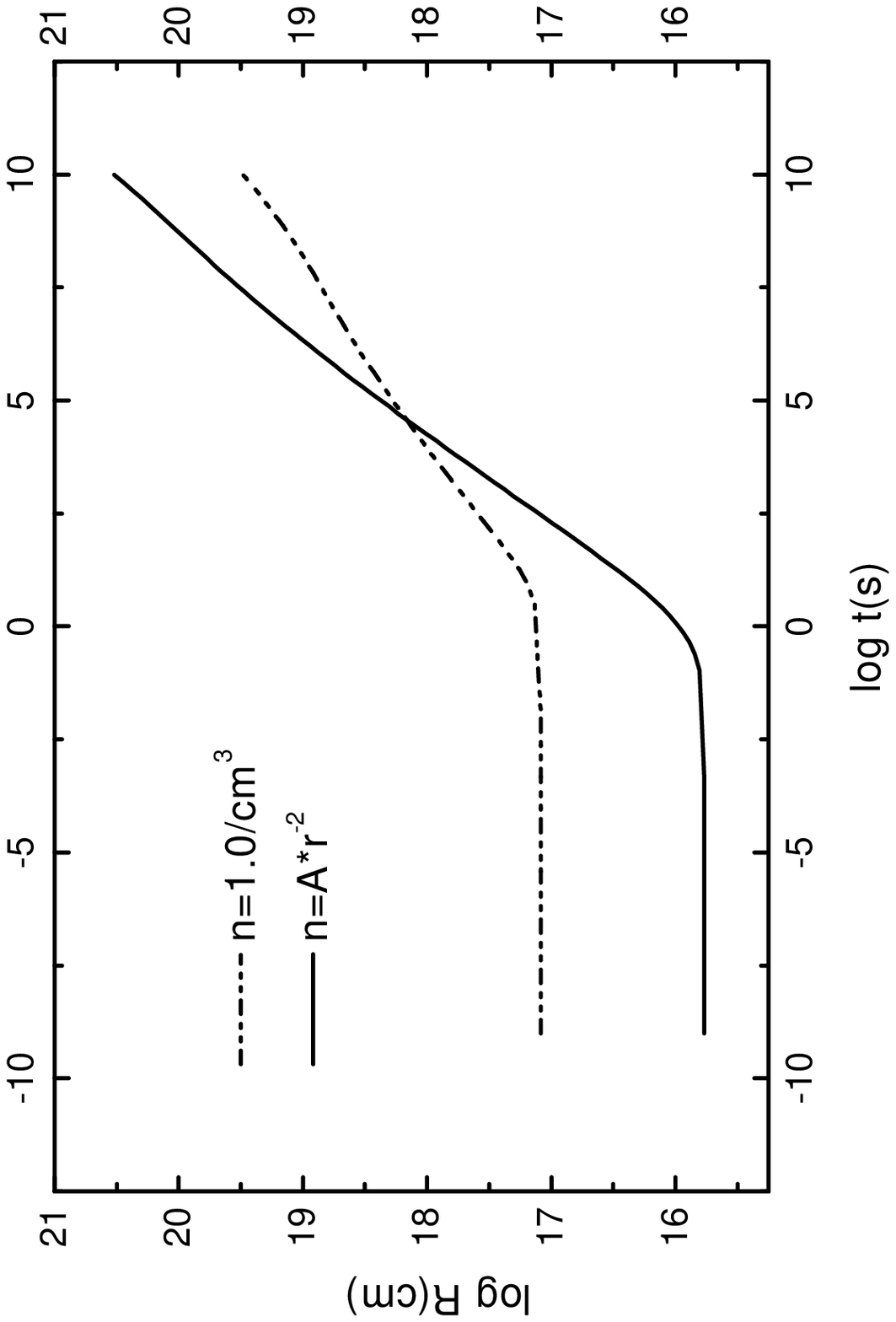}}
\end{picture}
\caption
{ Evolution of the shock radius ($R$).  Parameters and line styles are the same as in 
  Fig.1. }
\end{figure}

\noindent
where 
\begin{equation}
\label{dneiiic19}
C_4=\frac{{p}-1}{\gamma_{\rm e,min}^{1-{p}}-\gamma_{\rm e,max}^{1-{ p}}} N_{\rm ele}.
\end{equation}
\end{description}

\subsection{Formulae of Synchrotron Spectrum}

In the co-moving frame, the synchrotron radiation power at frequency $\nu '$ from 
electrons is given by (Rybicki \& Lightman 1979)
\begin{equation}
\label{pnue20}
P'(\nu ') = \frac{\sqrt{3} e^3 B'}{m_{\rm e} c^2} 
	    \int_{\gamma_{\rm e,min}}^{\gamma_{\rm e,max}} 
	    \left( \frac{dN_{\rm e}'}{d\gamma_{\rm e}} \right)
	    F\left(\frac{\nu '}{\nu_{\rm c}'} \right) d\gamma_{\rm e},
\end{equation}
where $e$ is electron charge, 
$\nu_{\rm c}' = 3 \gamma_{\rm e}^2 e B' / (4 \pi m_{\rm e} c)$, and 
\begin{equation}
\label{fx21}
F(x) = x \int_{x}^{+ \infty} K_{5/3}(k) dk,
\end{equation}
with $K_{5/3}(k)$ being the Bessel function.
 We assume that this power is radiated isotropically in the comoving frame,
\begin{equation}
\label{dpdwp22}
\frac{d P'(\nu ')}{d \Omega '} = \frac{P'(\nu ')}{4 \pi}.
\end{equation}

Defining $\mu = \cos \theta$, we can derive the differential power 
in the observer's frame (Rybicki \& Lightman 1979; Huang et al. 2000b, d), 
\begin{equation}
\label{dpdw23}
\frac{d P(\nu)}{d \Omega} = \frac{1}{\gamma^3 (1 - \beta \mu)^3}
			    \frac{dP'(\nu ')}{d \Omega '}
			  = \frac{1}{\gamma^3 (1 - \beta \mu)^3}
			    \frac{P'(\nu ')}{4 \pi},
\end{equation}
\begin{equation}
\label{tnue24}
\nu = \frac{\nu '}{\gamma (1 - \mu \beta)}. 
\end{equation}
Here quantities with prime are measured in the comoving frame, and  quantities without prime are in the observer's frame.
Then the observed flux density at frequency $\nu$ at certain angle is

\begin{figure}
\begin{picture}(100,170)
\put(0,0){\includegraphics{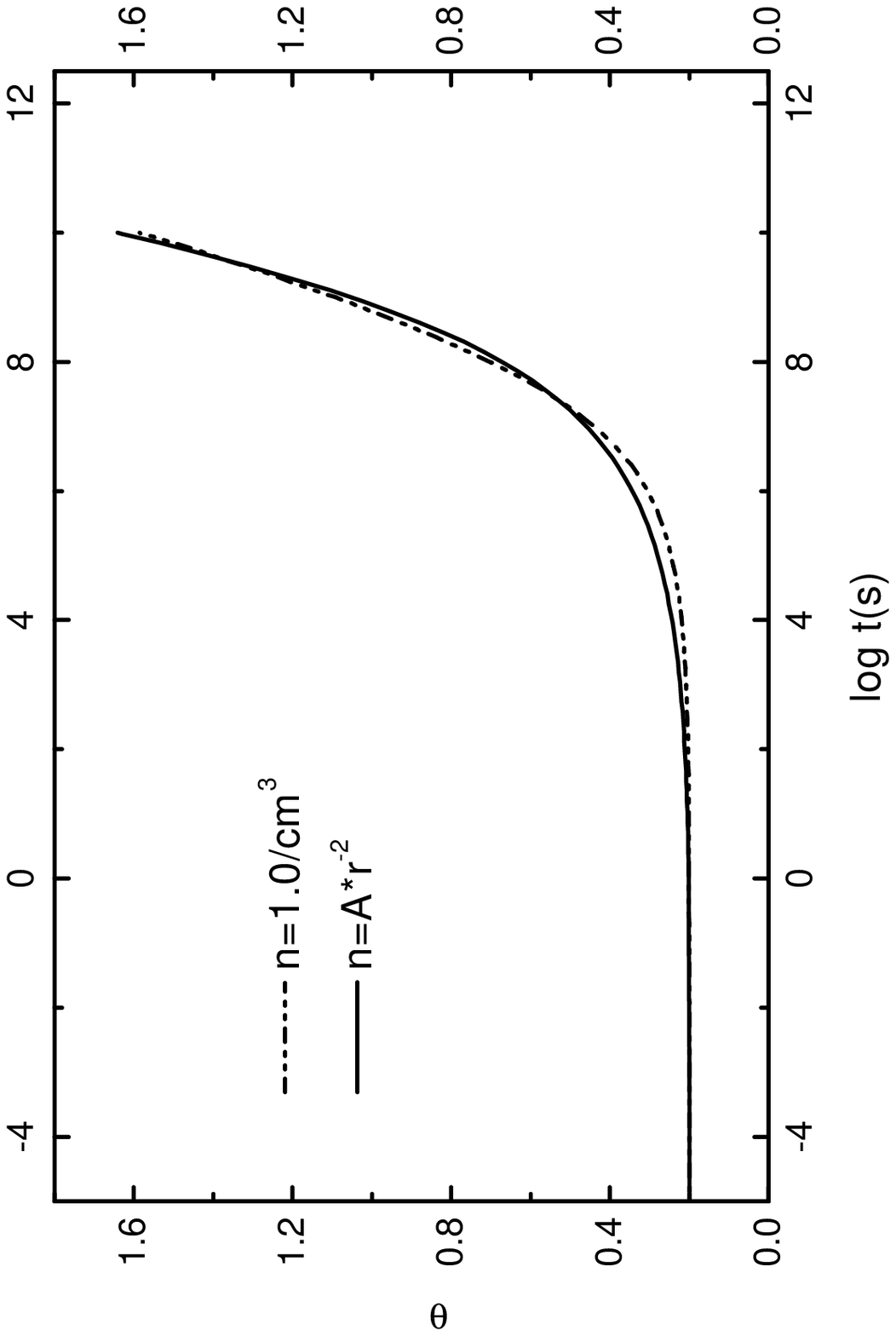}}
\end{picture}
\caption
{ Evolution of the half  opening angle($\theta$). Parameters and line styles are the same  as in Fig.1. }
\end{figure}

\begin{eqnarray}
\label{snue25}
s_{\nu}(\mu) &= & \frac{1}{A_{\rm s}} \left( \frac{dP(\nu)}{d \Omega} \frac{A_{\rm s}}{D_{\rm L}^2}\right) \nonumber \\
	& = & \frac{1}{\gamma^3 (1 - \beta \mu)^3} \frac{1}{4 \pi D_{\rm L}^2} P'\left(\gamma(1 - \mu \beta) \nu \right),
\end{eqnarray}

\noindent
where $A_{\rm s}$ is the area of the detector and $D_{\rm L}$ is the luminosity 
distance. The observed flux density at a given frequency is 
obtained by integrating over the shock front within the jet boundary $\theta_{\rm j}$.

\section{Numerical Results}

In our model, we use the following initial values or parameters as a set of ``standard'' parameters: $E_{\rm 0}/\Omega_{\rm 0}
=1 \times 10^{54}{\rm ergs}/4\pi, \gamma_{\rm 0} = 300$ (i.e. $M_{\rm ej}/ \Omega_{\rm 0} = 0.002M_\odot /4\pi$),  $\xi^{2}_{\rm B}=0.02$, $ p=2.5$,
$\xi_{\rm e}=0.1,  \theta_{\rm 0} = 0.2$. For simplicity, we 
assume that the expansion during the whole stage is adiabatic, i.e. $\epsilon \equiv 0$.

Figure 1 shows the evolution of the Lorentz factor. We see that the bulk Lorentz factor changes very slowly at late times in the 
wind case, compared with that in the homogeneous ISM case. For the wind case, using the standard parameters, we see that at 
$10^8$ s, the blast wave decelerates into the non-relativistic stage (here we let $\gamma=2$ be the critical point between 
relativistic and non-relativistic phases). In the homogeneous ISM case, the blast wave evolves into the 
non-relativistic  stage at about $10^7{\rm s}$.
 Figure 2 illustrates the time dependence of the shock radius.
 
\begin{figure}
\begin{picture}(100,170)
\put(0,0){\includegraphics{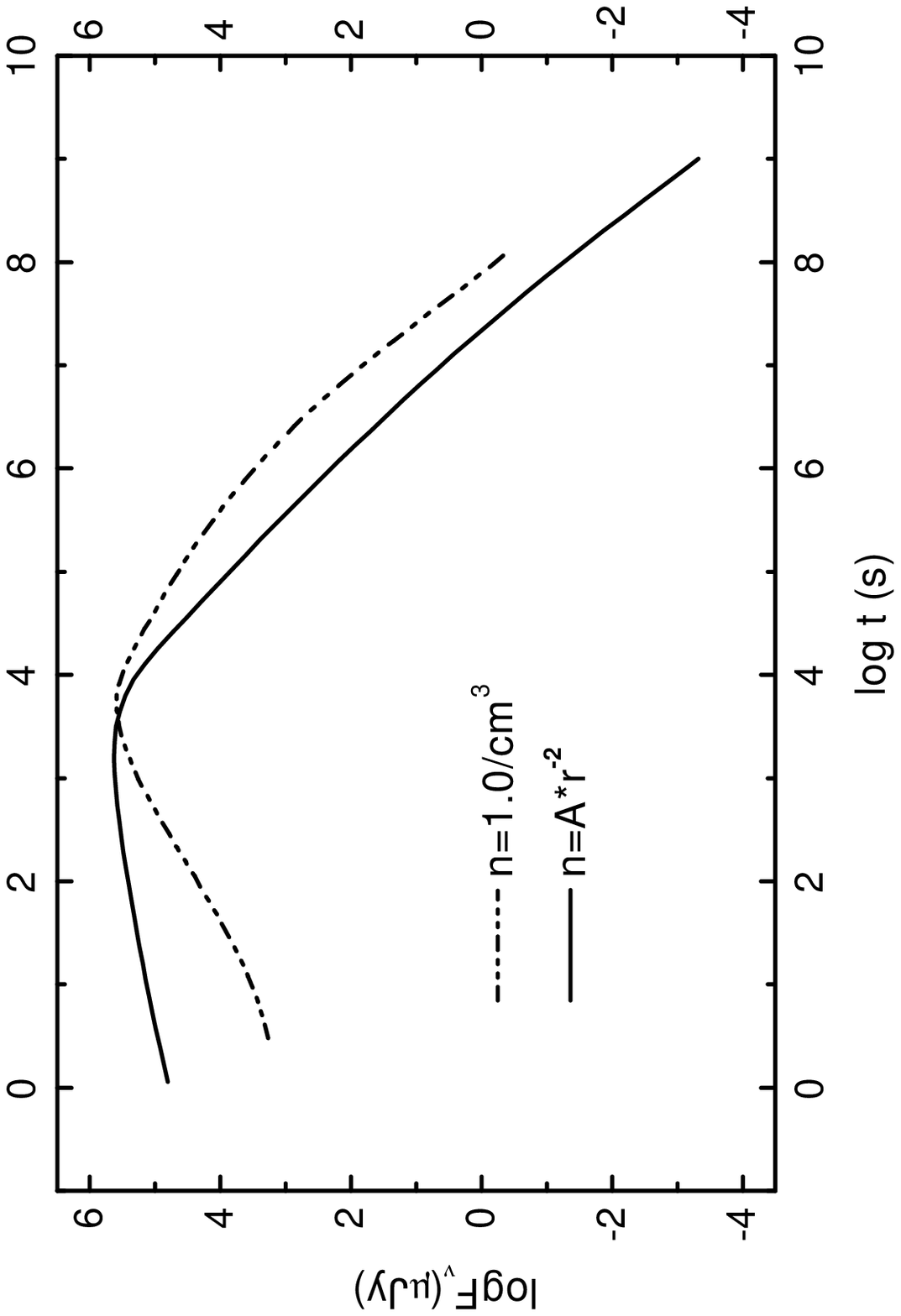}}
\end{picture}
\caption
{ R band afterglow light curves. The thick solid 
  line corresponds to a jet with ``standard'' parameters, and viewing angle $\theta_{\rm obs}$=0.} 
\end{figure}

In figure 3  we present the evolution of the jet opening angle. During the ultra-relativistic  phase the angle increases only 
slightly, because at this time, lateral expansion speed can be thoroughly neglected when compared with the blast 
wave speed itself. However at the Newtonian stage,   $\theta$ increases  quickly from 0.6 to 1.4.

In figure 4, we show the R-band light curve.  Please note that at the end point of each 
curve the average electron Lorentz factor is already as small as $\gamma_{\rm e,min}$ = 5, corresponding to a bulk Lorentz factor of 
$\gamma$ = 1.05, which implies that the jet is completely in the Newtonian regime.
Rhoads (1999) as well as Livio \& Waxman (1999) has predicted that the light curve will show a break when the bulk Lorentz factor 
is  $\gamma \approx 1/ \theta_{0}$ (here the blast wave is still at the relativistic stage ).
In figure 5, we give the evolution of the time index of the afterglow. We expect that the break in the wind case is less obvious than 
 that found by Huang et al. (2000a, b, c, d) in the homogeneous ISM. Our numerical results verify this expectation.

 Combining figure 4 with figure 1,  we can see that both in the wind environment and  in the homogeneous ISM environment 
 there is no observable break during the relativistic stage, which is consistent with the results of Panaitescu \& M\'{e}sz\'{a}ros (1998), 
Moderski et al. (2000), Huang et al. (2000a, b, c, d), and Wei \& Lu (2000).  We expected that there would be an obvious break during the trans-relativistic 
stage, i.e. transition from the relativistic stage to the non-relativistic stage. We don't find such a break but rather a smooth curve, which can be 
seen clearly in figure 5. In a uniform density medium the increase of the index in the power-law of the light curve is  1.07 during about 
two and a half decades in time. For a pre-ejected stellar wind $\alpha$ increases  by 0.8 over 5 decades. Therefore,  as argued by Kumar $\&$ 
Panaitescu (2000), a break in the light curve for a jet in a wind model is unlikely to be detected.

\section{Discussion}

Following Huang et al. (2000a, b, c, d), who considered the evolution of a jet in a homogeneous ISM environment, we investigated the detailed 
dynamical evolution of jets and their afterglows for the wind case from the ultra-relativistic stage to the non-relativistic stage. Recently, Kumar 
\& Panaitescu  (2000) considered the evolution of a jet in stratified media. Compared with their studies, our model is refined 
in the following aspects:
(1)	Kumar \& Panaitescu (2000) considered  the evolution of an adiabatic jet, while we study the expansion of a partially radiative
realistic jet. Furthermore, the dynamics presented are applicable to both ultra-relativistic and 
 Newtonian jets, so we could follow the overall evolution of a jet using a  set of differential equations. 
 (2) Similarly, Kumar \& Panaitescu (2000) did not consider the variation of sound velocity  with time.
We describe the lateral expansion of jets with a refined and more reasonable sound speed expression, which varies 
with the bulk Lorentz factor (Huang et al. 2000b, d). 

\begin{figure}
\begin{picture}(100,170)
\put(0,0){\includegraphics{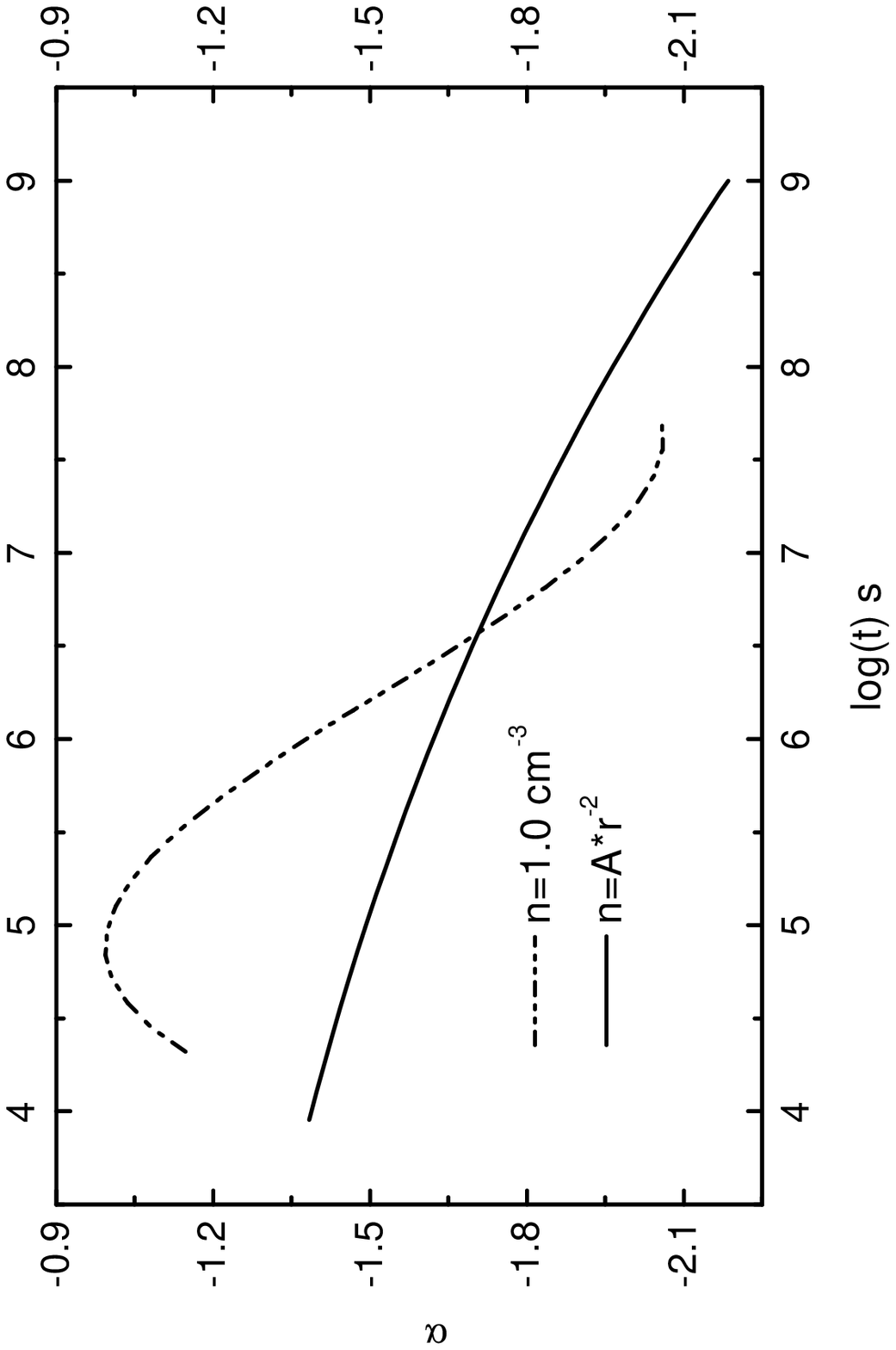}}
\end{picture}
\caption
{ The evolution of the time index of the afterglow light curve  at the R band. Parameters and line styles are the same as in  Fig.1.}
\end{figure}

In addition, we considered the evolution of electron distribution
 with time. Despite these differences, one of our results is similar to that of Kumar \& Panaitescu (2000): 
there seems no observable break around the time of $\gamma =1/\theta_{\rm 0}$, which  conflicts
 with Rhoads' (1999) and Livio \& Waxman's (1999) expectations.
 Furthermore, we also find that:

1.   Livio \& Waxman (1999) predicted that a light curve flattening would occur when the blast wave evolves into the non-relativistic 
stage. For the wind case, we calculated up to $10^9{\rm s}$ (corresponding to a Lorentz factor of 1.05), when the  blast wave has
 completely evolved into the non-relativistic stage. We do not find any flattening tendency in the light curve.

2.  The transition from the ultra-relativistic phase to the non-relativistic phase is also very smooth; the expected obvious 
break does not appear. This is very similar to the behaviour of isotropic fireballs (Wijers, Rees, \& M\'{e}sz\'{a}ros1997, Huang et al. 2000b).

3.  If we use the same parameters (except for the differences in number density for the wind case and for the 
homogeneous ISM), we find that  the flux 
density in the wind case is obviously weaker than that in the homogeneous ISM case. This property is consistent with  Chevalier \& Li's (2000)
result.

Two currently popular models for GRB progenitors are the mergers of compact objects (neutron stars or black holes) and  the 
explosions of massive stars. It is widely believed that GRBs produced by the former model occur in the ISM with density 
n $\sim 1 {\rm cm}^{-3}$ and  GRBs produced by the latter model occur in the preburst stellar wind enviroment with mass 
density $\rho \propto R^{-2}$. As argued by Chevalier \& Li (2000) and Livio \& Waxman (1999), both ISM and wind cases should show the same 
emission feature during the lateral spreading phase, and in particular on a timescale of days, the wind density is similar to 
typical ISM densities so that an interaction with the wind would give results that  are not different from the ISM case.
If GRBs are beamed, thus, their optical afterglow emission could not be used to discriminate the massive prognitor model 
from the compact binary progenitor model. However, our numerical results show that their optical afterglow emissions are 
 different, particularly several days after the burst. Thus, it may be used to discriminate the two models from each
other, and further observations may verify our numerical results.

\acknowledgements 

We are very grateful  to  Z. Li, Z.X. Ma and X.Y. Wang for useful  comments.  
This work was partially supported by the National Natural Science 
Foundation of China (grants 19773007, 19973003 and 19825109), and the National Project of Fundmental Researches (973 Project).

\end{document}